%% file: article.tex
\documentclass[12pt]{article}
\usepackage{epsfig }
\usepackage{lineno}
\input econfmacros.tex

\input babarsym
\input note2463-symbols

\def\Title#1{\begin{center} {\Large {\bf #1} } \end{center}}

\begin{document}
  \linenumbers

\Title{Combination of $\gamma$ measurements from BaBar}

\bigskip\bigskip


\begin{raggedright}  

{\it Denis Derkach\index{Derkach, D.}\\
INFN, Sezione di Bologna, I-40127 Bologna, ITALY\\
and\\
LAL, Orsay, F-91898, FRANCE\\
On behalf of the \babar\ collaboration}
\bigskip\bigskip
\end{raggedright}

Proceedings of CKM 2012, the 7th International Workshop on the CKM Unitarity Triangle, University of Cincinnati, USA, 28 September - 2 October 2012

\begin{abstract}
\noindent
We present the combination of the CKM angle $\gamma$ measurements performed by the \babar\ experiment at the
\pep2 \epem collider at SLAC National Laboratory. The analysis supersedes previous results obtained by collaboration
and gives $\gamma=\left(69^{+17}_{-16}\right)^{\circ}$ modulo $180^{\circ}$. The results are inconsistent with the absence of \CP\ violation at a significance of 5.9 standard deviations.
\end{abstract}

\section{Introduction}

The Cabibbo-Kobayashi-Maskawa (CKM)~\cite{cite:CKM} angle \g\ is one of the least 
precisely known parameters of the unitarity triangle. 

Several methods have been proposed to extract \g. 
Those using charged \B meson decays into $D^{(*)}K^{(*)}$ final states 
have no penguin contribution, which gives an important difference from most of other direct measurements of the angles. 
These processes are theoretically clean provided that hadronic unknowns are determined from experiment. 
The $b\to c\ubar s$ and
$b \to u \cbar s$ tree amplitudes are used to construct the observables that depend
on their relative weak phase \g, on 
the magnitude ratio $\rb \equiv | {\cal A}(b\to u\cbar s) / {\cal A}(b\to c\ubar s) |$
and on the relative strong phase $\deltab$ between the two amplitudes.

The various methods can be classified by the neutral $D$ decay final state that is reconstructed~\cite{cite:charge}. The three main approaches 
employed by the $B$ factory experiments are:
\begin{itemize}
\item the Dalitz plot (DP) or Giri-Grossman-Soffer-Zupan (GGSZ) method, based on 3-body, self-conjugate final states, such as $\KS\pi\pi$~\cite{cite:dalitz_theo};
\item the Gronau-London-Wyler (GLW) method, based on decays to CP eigenstates, such as $\Kp\Km$ and $\KS\piz$~\cite{cite:glw_theo};
\item the Atwood-Dunitz-Soni (ADS) method, based on $D$ decays to doubly-Cabibbo-suppressed final states, such as $\Dz \to K\pi$~\cite{cite:ads_theo}.
\end{itemize}

The \babar\ collaboration that analyzes data recorded at the asymmetric \epem
collider \pep2 at SLAC national laboratory, have produced several important results in the field. These results can be combined into 
a single number using all available information including the experimental information, which was not previously published by \babar.  
In the following, we show the results of the combination of GGSZ~\cite{cite:GGSZ2010}, GLW~\cite{cite:GLW_d0k,cite:GLW_dstar0k,cite:GLWADS_d0kstar}, and
ADS~\cite{cite:ADS_d0k_dstar0k_kpi,cite:ADS_d0k_kpipi0} analyses, performed by \babar. These analyses are based on 474 millions \BB pairs at most.
Results from Belle and LHCb were presented at this conference too. A more complete discussion of analysis can be seen at~\cite{Lees:2013zd}.

\section{Combination Method}

\begin{table}[htb] 
\begin{center} 
\begin{tabular}{l|cc} 
           & Real part (\%)&  Imaginary part (\%)\\ \hline  
 $\zbm$    & $\phm8.1\pm2.3\pm0.7$    	     & $\phm4.4\pm3.4\pm0.5$ \\ 
 $\zbp$    & $-9.3\pm2.2\pm0.3$          	     & $-1.7\pm4.6\pm0.4$ \\ 
 $\zbstm$  & $-7.0\pm3.6\pm1.1$                & $-10.6\pm5.4\pm2.0$    \\ 
 $\zbstp$  & $\phm10.3\pm2.9\pm0.8$             & $-1.4\pm8.3\pm2.5$ \\ 
 $\zsm$    & $\phm13.3\pm8.1\pm2.6$    	     & $\phm13.9\pm8.8\pm3.6$    \\ 
 $\zsp$    & $-9.8\pm 6.9\pm1.2$               & $\phm11.0\pm 11.0\pm6.1$    \\ 
\end{tabular} 
\caption{\label{tab:xyresults} 
\CP-violating complex parameters  
$\zbzbstpm = \xbxbstpm + i \ybybstpm$ 
and  
$\zspm = \xspm + i \yspm$ 
obtained from the combination of GGSZ, GLW, and ADS measurements.
The first error is statistical (corresponding from $-\twoDLL=1$), 
the second is the experimental systematic uncertainty including  
the systematic uncertainty associated to the GGSZ decay amplitude models.
} 
\end{center} 
\end{table} 
We combine all the GGSZ, GLW, and ADS observables ($34$ in total) to extract \g in two different stages. 
First, we extract the best-fit values for the \CP-violating quantities in terms of the GGSZ analysis observables given as
\begin{equation}
\zbzbstpm=\rbrbstpm e^{i(\deltabdeltabst \pm \g)}
\end{equation}
and
\begin{equation} 
\zspm = \kappa \rspm e^{i(\deltas \pm \g)},
\end{equation}
for $B^\pm\to D^{(*)}K^{\pm}$ and $B^\pm\to D K^{*\pm}$ decays,
respectively.
The hadronic parameter $\kappa$ is defined as
\begin{equation} 
\kappa e^{i\delta_{s}} \equiv \frac{\int A_c(p) A_u(p) e^{i\delta(p)}{\rm d}p}{\sqrt{\int A^2_c(p){\rm d}p \int A^2_u(p){\rm d}p}},
\label{eq:kappa}
\end{equation}
where $A_c(p)$ and $A_u(p)$ are the magnitudes of the $b \to c \ubar s$ and $b\to u \cbar s$ amplitudes as a function of 
the $\Bpm \to \D \KS \pipm$ phase space position $p$, and $\delta(p)$ is their relative strong phase. 
This coherence factor, with $0<\kappa<1$ in the most general case and $\kappa=1$ for two-body \B decays,
accounts for the interference between $\Bpm \to\D \Kstarpm$ and other $\Bpm \to\D \KS\pi^\pm$ decays,
as a consequence of the \Kstarpm natural width~\cite{ref:gronau2003}.
In our analysis $\kappa$ has been fixed to $0.9$~\cite{cite:GGSZ2010}, and a 
systematic uncertainty has been assigned varying its value by $\pm0.1$.
Thus the parameter $\delta_{s}$ is an effective strong-phase 
difference averaged over the phase space.
\begin{table}[htb]
\begin{center}
\begin{tabular}{l|cc} 
Parameter            & {$68.3\%$ \CL}  & {$95.5\%$ \CL} \\ 
\hline \noalign{\vskip3pt}
\g $(^\circ)$         & $69^{+17}_{-16}$      & $[41,102]$    \\
\rb $(\%)$            & $9.2 ^{+1.3}_{-1.2}$  & $[6.0,12.6]$  \\
\rbst $(\%)$          & $10.6^{+1.9}_{-3.6}$  & $[3.0,14.7]$  \\
\krs $(\%)$           & $14.3^{+4.8}_{-4.9}$  & $[3.3,25.1]$  \\
\deltab   $(^\circ)$  & $105^{+16}_{-17}$     & $[72,139]$    \\
\deltabst $(^\circ)$  & $-66^{+21}_{-31}$     & $[-132,-26]$  \\
\deltas   $(^\circ)$  & $101\pm43$            & $[32,166]$    \\
 
\end{tabular}
\caption{\label{tab:polarresults} $68.3\%$ and $95.5\%$ 1-dimensional \CL regions,
equivalent to one- and two-standard-deviation intervals,
for \g, \deltabdeltabst, \deltas, \rbrbst, and \krs,
including all sources of uncertainty, obtained from the combination of GGSZ, GLW and ADS measurements.
The results for \g, \deltabdeltabst\ and \deltas\ are given modulo a $180^\circ$ phase.
}
\end{center}
\end{table}

\begin{figure}[htb!]
\begin{center}
\epsfig{file=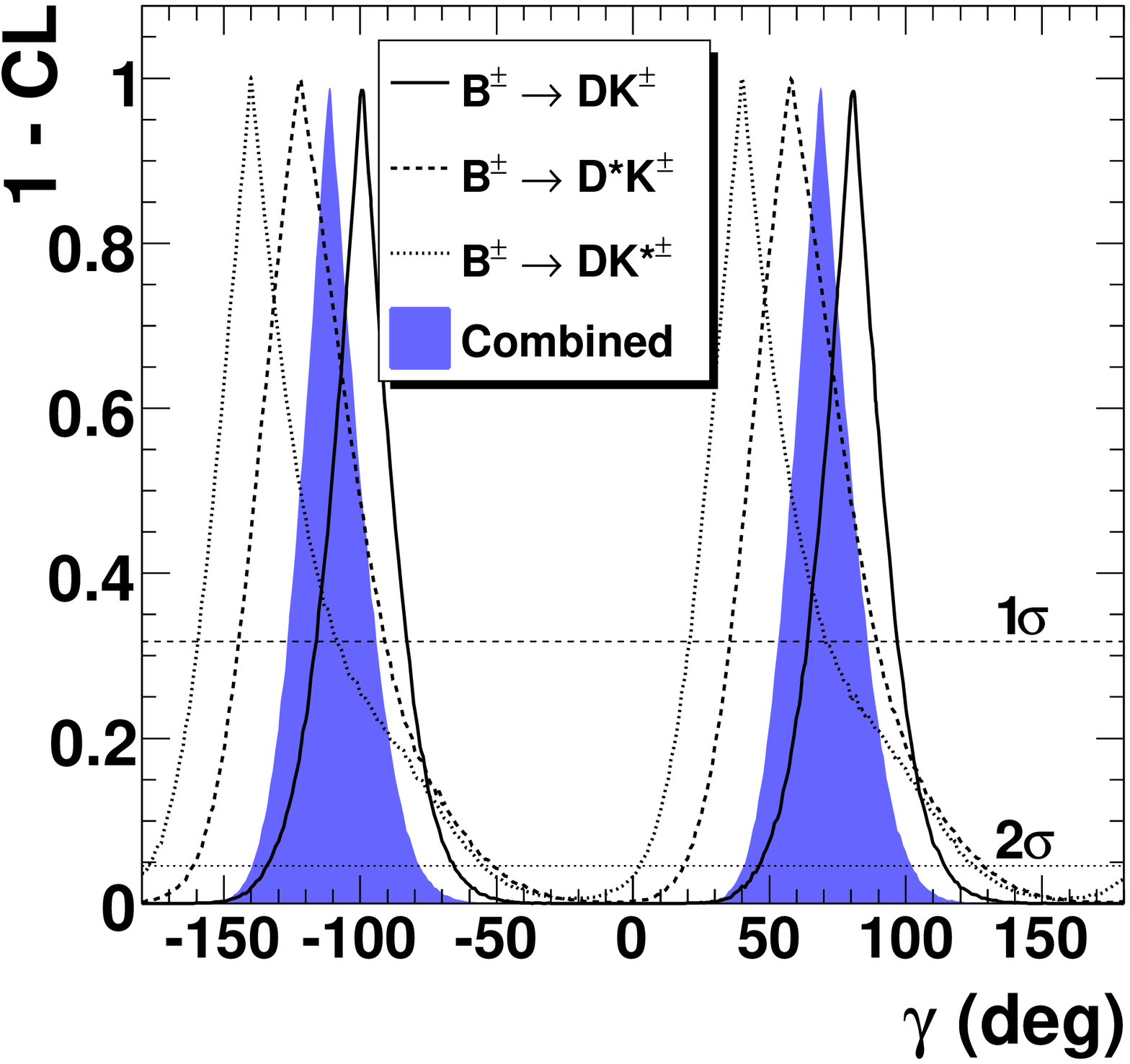,width=0.3\linewidth}
\epsfig{file=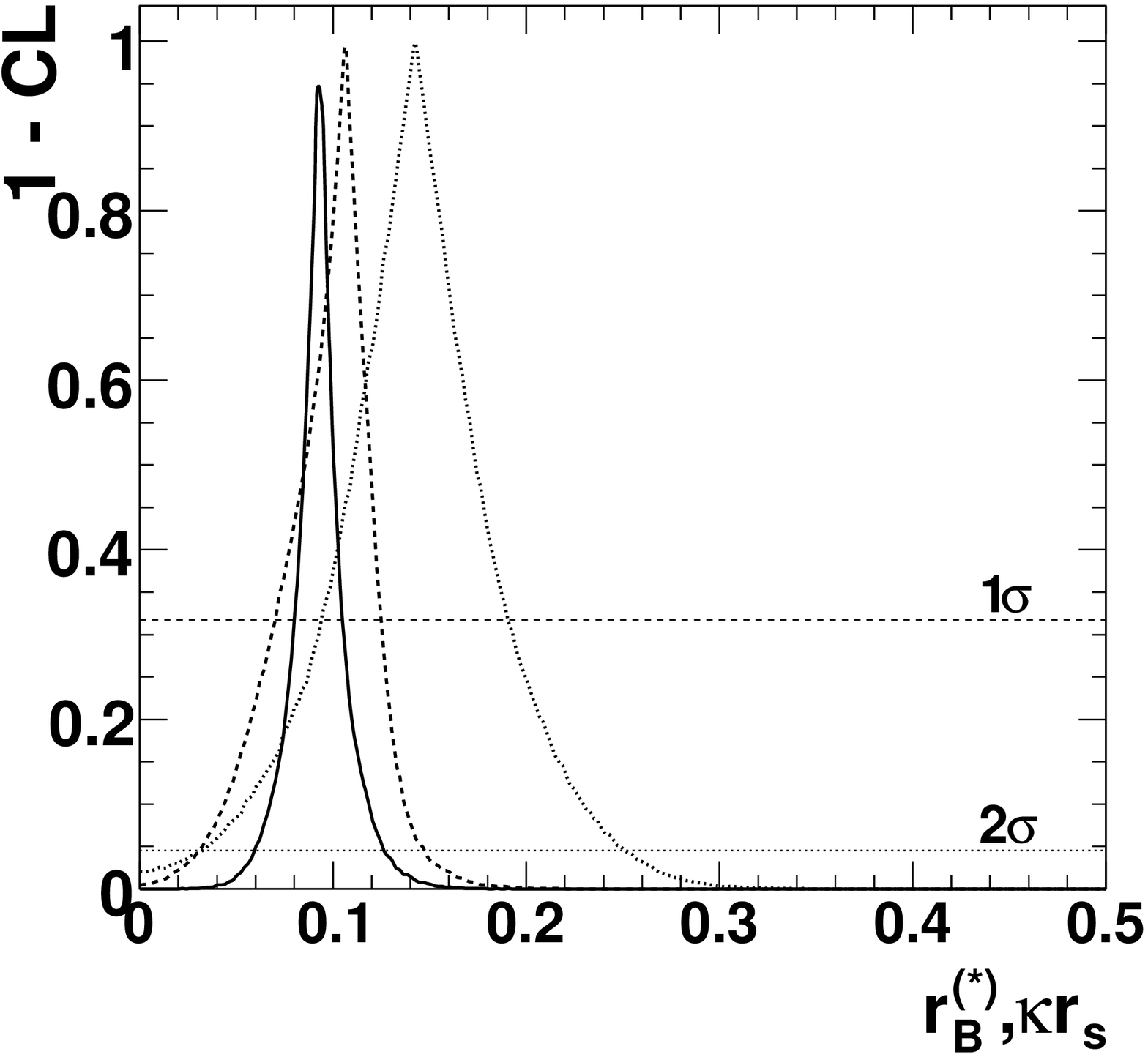,width=0.3\linewidth}
\epsfig{file=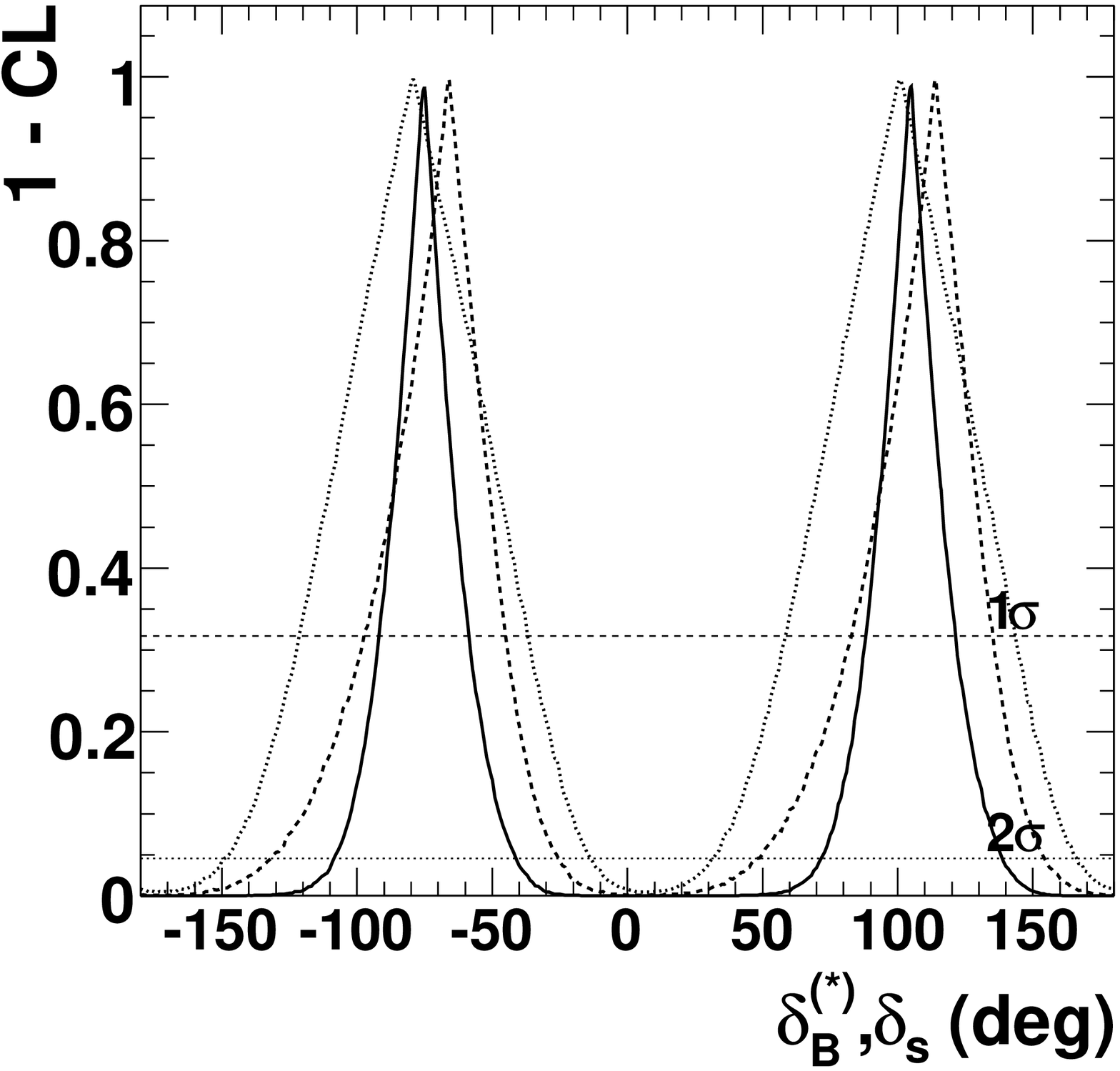,width=0.3\linewidth}
\end{center}
\caption{(color online). Combined $1-\CL$ as a function of \g (left), \rbrbst, and \krs (middle), and \deltabdeltabst, \deltas (right),
including statistical and systematic uncertainties,
for $\Bpm \to \D \Kpm$, $\Bpm \to \Dstar \Kpm$, and $\Bpm \to \D \Kstarpm$ decays. The combination of all the \B decay channels is also shown for \g.
The dashed (dotted) horizontal line corresponds to the one- (two-) standard-deviation \CL.
\label{fig:scans-gamma-rb-delta}
}
\end{figure}
\begin{figure}[htb!]
\begin{center}
\begin{tabular} {c}
\epsfig{file=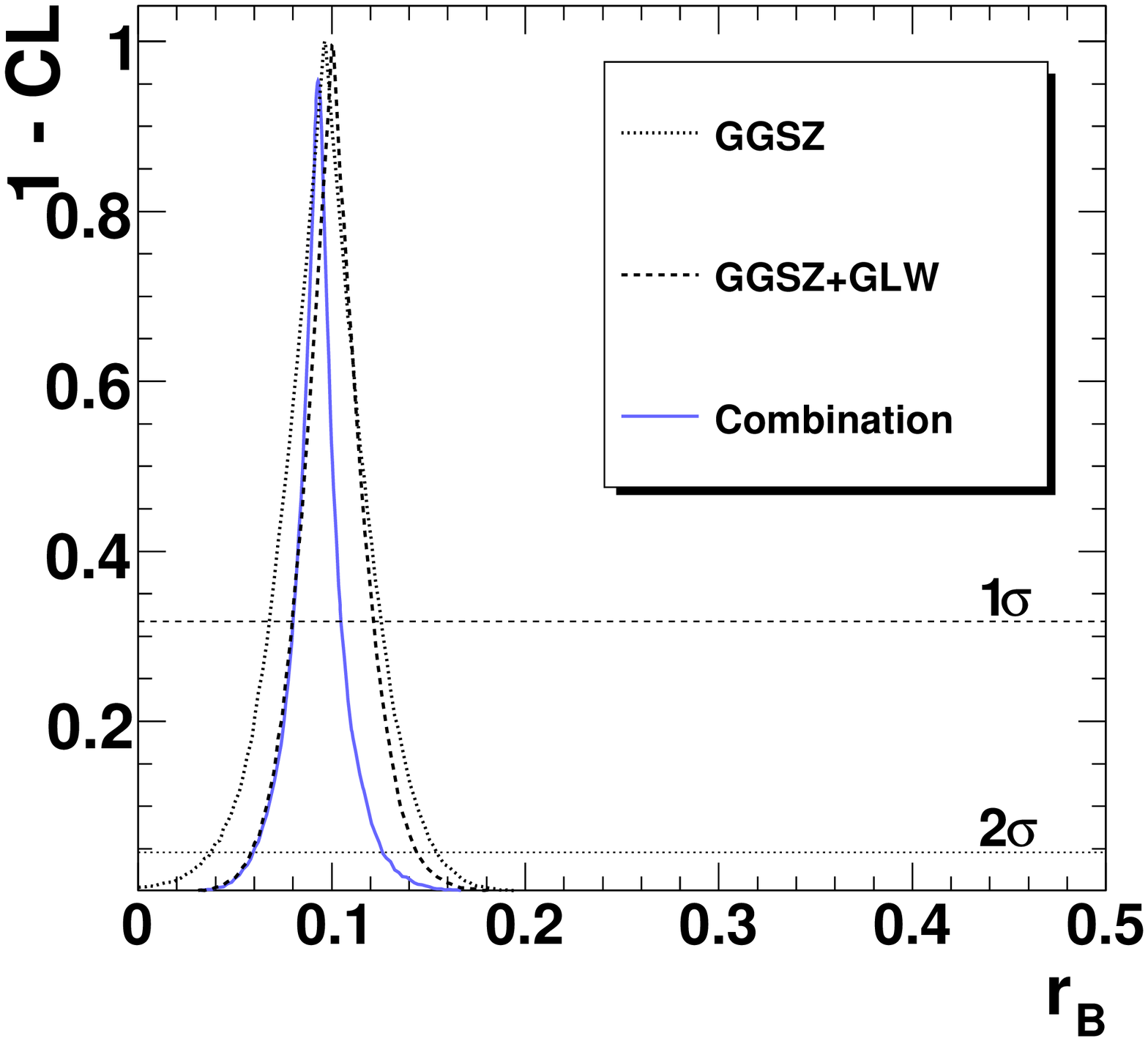,width=0.3\linewidth}
\epsfig{file=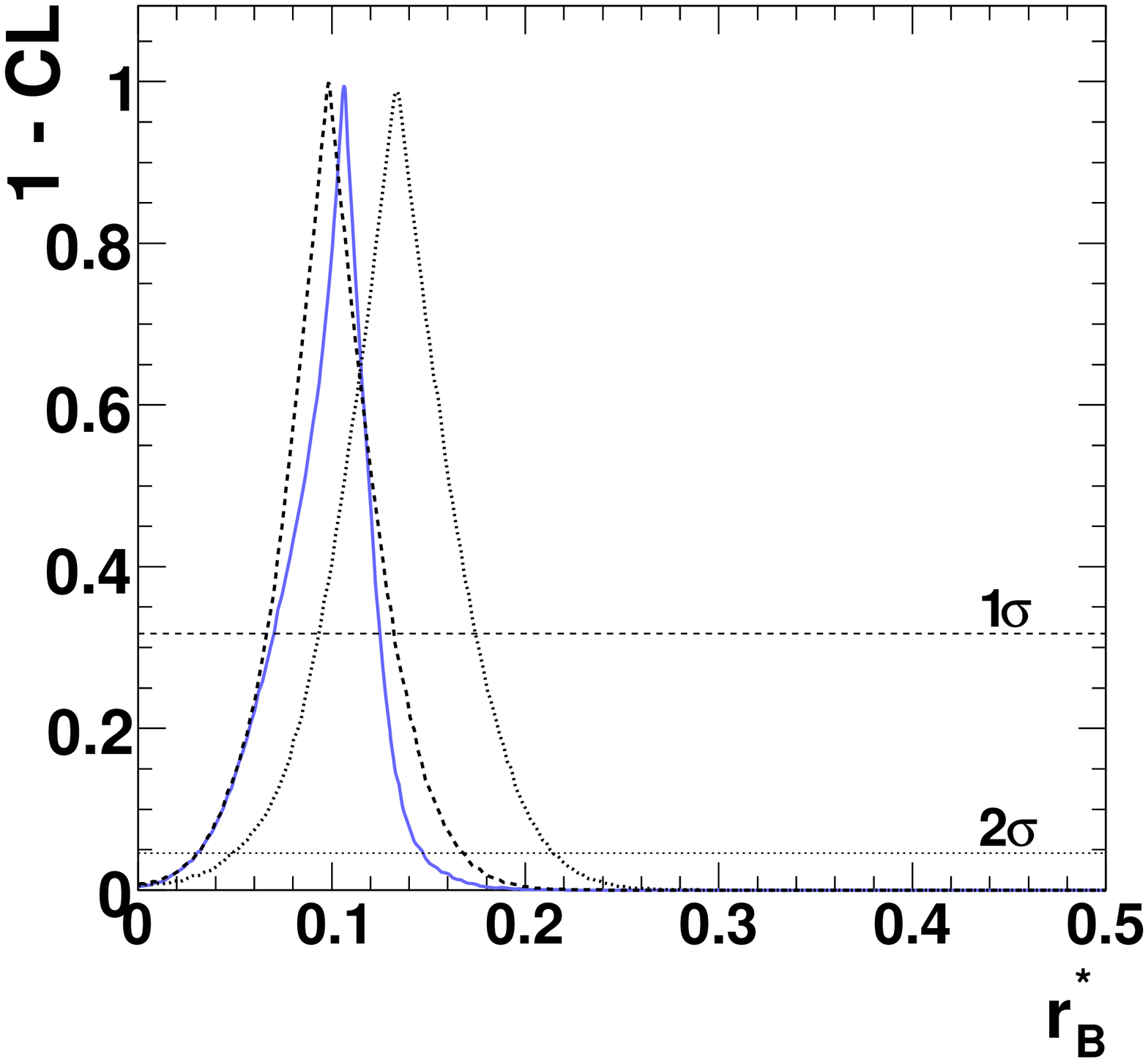,width=0.3\linewidth} 
\epsfig{file=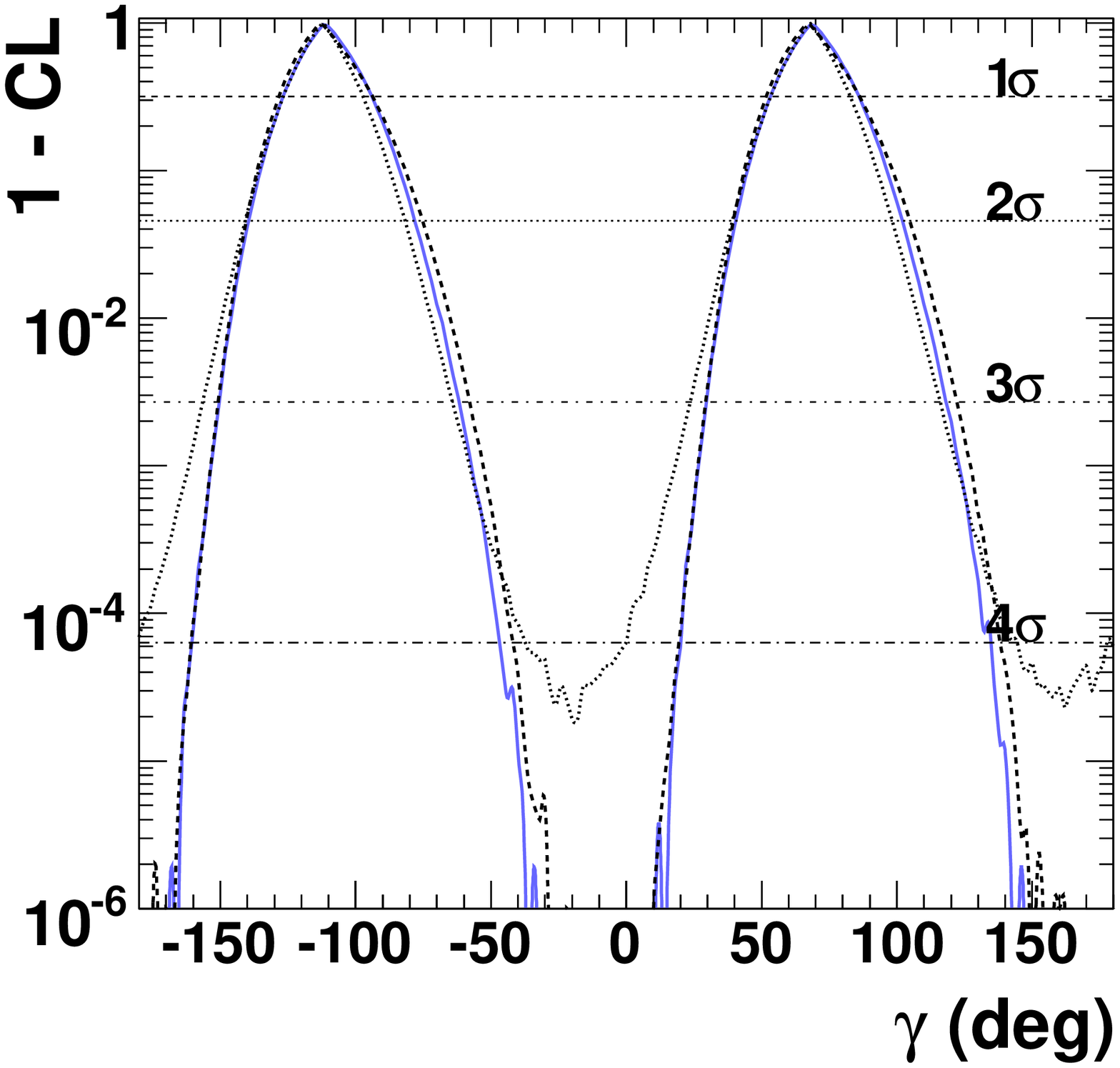,width=0.3\linewidth} 
\end{tabular}
\end{center}
\caption{(color online). Comparison of $1-\CL$ as a function of \rb (left), \rbst (middle), and \g (right) for all \B decay channels combined with the 
GGSZ-only method only, the combination with the GLW measurements, and the global combination, 
including statistical and systematic uncertainties.
The horizontal lines represent the one-, two-, three- and four-standard-deviation \CL.
\label{fig:scan-comp-rb+gamma}
}
\end{figure}
The combination also profits from external inputs for the \D hadronic parameters: amplitudes ratio $r_D$, 
strong phase $\delta_D$, and coherence factor $k_D$. These
are taken from PDG~\cite{ref:pdg2010} and CLEO-c~\cite{ref:cleoKpipi0} results.
All external observables are assumed to be
uncorrelated with the rest of the input observables, while we take into account the
correlation measured by CLEO-c. 

The best-fit values of \zbzbstpm and \zspm are obtained by maximizing a
combined likelihood function constructed as the product of partial 
likelihood \pdfs for GGSZ, GLW, and ADS measurements.
For the decays $\Bpm\to \Dstar_{\CP-}[\D_{\CP-}\piz] \Kpm$, 
$\Bpm\to \Dstar_{\CP+}[\D_{\CP-}\gamma] \Kpm$, and $\Bpm\to \D_{\CP-} \Kstarpm$, measurement without the $D\to\KS\phi$ 
channel, which is common in GGSZ and GLW analyses is not available.
The impact is estimated by increasing the 
uncertainties quoted in Refs.~\cite{cite:GLW_dstar0k,cite:GLWADS_d0kstar}
by 10\% while keeping the central values unchanged. This is done in accordance 
to the study performed in the $\Bpm\to \D_{\CP-} \Kpm$ analysis~\cite{cite:GLW_d0k}.
The results for the combined 
\CP-violating parameters \zbzbstpm and \zspm
are summarized in Table~\ref{tab:xyresults}.

In a second step, we transform the
measurements from Table~\ref{tab:xyresults} into 
the physically relevant quantities \g and the set of hadronic parameters
$\uvec \equiv (\rb, \rbst, \krs, \deltab, \deltabst, \deltas)$.
We adopt a
frequentist procedure~\cite{ref:pdg2010} to obtain one-dimensional confidence intervals
of well defined \CL that takes into account
non-Gaussian effects due to the nonlinearity of the relations between the observables and physical quantities.
Figure~\ref{fig:scans-gamma-rb-delta} illustrates $1-\CL$ as a function of \g, \rbrbst, \krs, \deltabdeltabst, and \deltas,
for each of the three \B decay channels separately and, in the case of \g, their combination.
From these distributions we extract one- and two-standard-deviation intervals as the sets of values for which
$1-\CL$ is greater than 31.73\% and 4.55\%, respectively, as summarized in Table~\ref{tab:polarresults}.
To assess the impact of the GLW and ADS observables in the determination of \g, we compare $1-\CL$ as a function of \rbrbst and \g 
for all \B decay channels combined using the GGSZ method alone, the combination with the GLW measurements, and the global combination, 
as shown in Fig.~\ref{fig:scan-comp-rb+gamma}. While the constraints on \rb are clearly 
improved at one- and two-standard-deviation level, and to a lesser extent on \rbst, their best (central) values move towards slightly
lower values. Since the uncertainty on \g scales roughly as $1/\rbrbst$,
the constraints on \g at $68.3\%$ and $95.4\%$ \CL do not improve compared to the GGSZ-only results,
in spite of the tighter constraints on the combined measurements shown in Table~\ref{tab:xyresults}.
However, adding GLW and ADS information reduces the confidence intervals for smaller $1-\CL$, as a consequence of
the more Gaussian behavior when the significance of excluding $\rbrbst=0$ increases.

The significance of direct \CP violation is obtained by evaluating $1-\CL$ for the most probable \CP conserving 
point, {\em i.e.}, the set of hadronic parameters \uvec with $\g=0$. Including 
statistical and systematic uncertainties, we obtain $1-\CL=3.4\times 10^{-7},\ 2.5\times 10^{-3}$, and $3.6\times 10^{-2}$,
corresponding to $5.1$, $3.0$, and $2.1$ standard deviations, for
$\Bpm\to\D\Kpm$, $\Bpm\to\Dstar\Kpm$, and $\Bpm\to\D\Kstarpm$ decays, respectively.
For the combination of the three decay modes we obtain $1-\CL=3.1\times 10^{-9}$, corresponding to $5.9$ standard deviations.

\section{Conclusions}

We determine $\g =
(69^{+17}_{-16})^\circ$ (modulo $180^\circ$), where the total 
uncertainty is dominated by the statistical component, with 
the experimental and amplitude model systematic uncertainties
amounting to $\pm 4^\circ$.

The combined significance of $\g \ne 0$ is $1-\CL=3.1\times 10^{-9}$,
corresponding to $5.9$ standard deviations, meaning
observation of direct \CP violation in the measurement of \g.

\end{document}

%% file: econfmacros.tex
\def\beq{\begin{equation}}
\def\eeq#1{\label{#1}\end{equation}}
\def\eeqn{\end{equation}}

\def\beqa{\begin{eqnarray}}
\def\eeqa#1{\label{#1}\end{eqnarray}}
\def\eeqan{\end{eqnarray}}

\let\bar=\overbar

\def\D{{\cal D}}

\def\Dslash{\not{\hbox{\kern-4pt $D$}}}
\def\dslash{\not{\hbox{\kern-2pt $\del$}}}

\def\BR{\mbox{\rm BR}}

\def\msb{{\bar{\ssstyle M \kern -1pt S}}}



%% file: note2463-symbols.tex
\def\babar{\mbox{\slshape B\kern-0.1em{\smaller A}\kern-0.1em
    B\kern-0.1em{\smaller A\kern-0.2em R}}\xspace}

\def\CL {\ensuremath{ \rm C.L. }\xspace}
\def\twoDLL  {\ensuremath{2\Delta\ln{\cal L}}\xspace}

\def \uvec {\ensuremath {{\mathbf u}}\xspace}

\def \pdfs {{P.D.F.s}\xspace}

\def\beq{\begin{equation}}
\def\eeq{\end{equation}}
\def\bea{\begin{eqnarray}}
\def\eea{\end{eqnarray}}
\def\bq{\begin{quote}}
\def\eq{\end{quote}}
\def\ben{\begin{enumerate}}
\def\een{\end{enumerate}}

\newcommand{\phm}{\ensuremath{\phantom{-}}}

\def\D {\ensuremath{D}\xspace}

\def \rb {\ensuremath {r_\B}\xspace}

\def \rbst {\ensuremath {r^\ast_\B}\xspace}
\def \rbrbst {\ensuremath {r^{(\ast)}_\B}\xspace}
\def \rbrbstpm {\ensuremath {r^{(\ast)}_{\Bpm}}\xspace}

\def \rs {\ensuremath {r_s}\xspace}
\def \rspm {\ensuremath {r_{s\pm}}\xspace}

\def \krs {\ensuremath {\kappa \rs}\xspace}
\def \deltab {\ensuremath {\delta_\B}\xspace}
\def \deltabst {\ensuremath {\delta^\ast_\B}\xspace}
\def \deltabdeltabst {\ensuremath {\delta^{(\ast)}_\B}\xspace}
\def \deltas {\ensuremath {\delta_s}\xspace}

\def \zbp {\ensuremath {{\mathsf z}_+}\xspace}
\def \zbm {\ensuremath {{\mathsf z}_-}\xspace}

\def \zbstp {\ensuremath {{\mathsf z}_+^\ast}\xspace}
\def \zbstm {\ensuremath {{\mathsf z}_-^\ast}\xspace}

\def \xbxbstpm {\ensuremath {{x}_\pm^{(\ast)}}\xspace}

\def \ybybstpm {\ensuremath {{y}_\pm^{(\ast)}}\xspace}

\def \zbzbstpm {\ensuremath {{\mathsf z}_\pm^{(\ast)}}\xspace}

\def \xspm {\ensuremath {{x}_{s\pm}}\xspace}

\def \yspm {\ensuremath {{y}_{s\pm}}\xspace}

\def \zsp {\ensuremath {{\mathsf z}_{s+}}\xspace}
\def \zsm {\ensuremath {{\mathsf z}_{s-}}\xspace}
\def \zspm {\ensuremath {{\mathsf z}_{s\pm}}\xspace}

\def\figurebox#1#2#3{%
    \def\arg{#3}%
    \ifx\arg\empty
    {\hfill\vbox{\hsize#2\hrule\hbox to #2{\vrule\hfill\vbox to #1{\hsize#2\vfill}\vrule}\hrule}\hfill}%
    \else
    {\hfill\epsfbox{#3}\hfill}%
    \fi}


%% file: article.bbl
\begin{thebibliography}{99}
\bibitem{cite:charge} Charge conjugate modes are implicitly included unless otherwise stated.
\end{thebibliography}
